\documentclass[journal,draftcls,onecolumn,12pt,twoside]{IEEEtran}

\ifCLASSINFOpdf
\else
\fi
\usepackage{amssymb}
\usepackage{amsmath,bm}
\usepackage{algorithm}
\usepackage{cite}
\usepackage{mathtools}
\usepackage{array}
\usepackage{multicol}
\usepackage{multirow}
\usepackage{graphicx}
\usepackage{epstopdf}
\usepackage[top=2.5cm, bottom=2.5cm, left=1.5cm, right=1.5cm]{geometry}
\usepackage{caption}
\usepackage{subfigure}
\usepackage{setspace}
\usepackage{longtable}
\renewcommand{\baselinestretch}{2}
\usepackage{booktabs}

\usepackage{enumitem}
\usepackage{tabu}
\usepackage{flushend}
\allowdisplaybreaks

\newcommand\eqa{\mathrel{\overset{\makebox[0pt]{\mbox{\normalfont\tiny\sffamily (a)}}}{=}}}
\newcommand\eqb{\mathrel{\overset{\makebox[0pt]{\mbox{\normalfont\tiny\sffamily (b)}}}{=}}}
\newcommand\eqc{\mathrel{\overset{\makebox[0pt]{\mbox{\normalfont\tiny\sffamily (c)}}}{=}}}

\newcommand\leqb{\stackrel{\mathclap{\normalfont\mbox{\normalfont\tiny\sffamily (b)}}}{\leq}}

\usepackage[utf8]{inputenc}

\begin{document}

%
\title{Performance Analysis of Energy Harvesting Underlay Cooperative Cognitive Radio Relay Networks with Randomly Located Nodes}

\author{Anupam~Shome,
        Amit~Kumar~Dutta,
        Saswat~Chakrabarti,
        Priyadip Ray
\thanks{A. Shome, A.K. Dutta, S. Chakrabarti are with G.S.Sanyal School of Telecommunications, IIT Kharagpur, India,
West Bengal, India, e-mail: (anupamshome06@yahoo.co.in, amitdutta@gssst.iitkgp.ac.in, saswat@ece.iitkgp.ernet.in).}
\thanks{Priyadip~Ray was with G.S.Sanyal School of Telecommunications, IIT Kharagpur, India,
West Bengal, India, e-mail: (priyadipr@gmail.com)}
      
        }


%


\maketitle

\begin{abstract}
In this work, we investigate the successful data communication probability of an energy harvesting co-operative cognitive radio network (CRN) in the presence of Poisson field of primary users (PU). We consider the scenario where, after harvesting energy from primary transmitters (PTs), the secondary transmitter (ST) would transmit its symbol towards secondary destination (SD) through a suitable secondary relay from group of randomly scattered  idle nodes within a circular region. We have considered several relay selection criteria in our work for a better relay node selection. We have also analytically evaluated the performance of secondary transmitter in terms of probability of successful symbol transmission. The relationship between the performance of ST and several network entities like density of PUs, transmit power of PTs and required transmit power of ST have been investigated through detailed analysis. The non-trivial trade-off between benefit of energy harvesting and interference from PTs has been explored in this present work. Numerical results are provided to verify the  precision of derived analytical expressions.
\end{abstract}


%

\section{ Introduction}
Wireless energy harvesting \cite{sudevalayam2011energy,mohjazi2015rf} provides a greener alternative to charge the batteries of sensor devices where frequent battery replacement of those devices is inconvenient and undesirable. Utilizing the concept of energy harvesting, sensor devices can charge their batteries from various cost-free sources of energy i.e solar, wind or RF signals without taking power from grid. In future, self-sufficient internet-of-things (IoT) devices are expected to be able to utilize the various energy sources to prolong their battery lifetime. \par
Over the past few years, issue of effective spectrum utilization has drawn considerable attention to the research community. With the enormous growth of various applications and wireless tele-taffic, efficient usage of spectrum has assumed significant importance. To alleviate the issue of effective spectrum usage, the concept of cognitive radio (CR) has been proposed~\cite{haykin2005cognitive}. Through the concept of CR, unlicensed or secondary users (SUs) can share the spectrum with licensed primary users (PUs), while causing little or no disturbance on primary data transmission. The SUs can access the spectrum allocated to PUs, following three approaches namely interweave~\cite{haykin2005cognitive}, underlay~\cite{goldsmith2009breaking}, overlay~\cite{sun2013overlay}. Further, through the process of energy harvesting, SUs can harness energy from various energy sources and they can charge their batteries without taking energy from power grid. Thus, it can be realized that the combination of energy harvesting and cognitive radio network can achieve green communication and efficient spectrum utilization both at the same time. \par
In the recent past, researchers have looked at several issues related to wireless energy harvesting considering non-cooperative as well as cooperative networks. In \cite{lee2013opportunistic}, considering stochastic geometry framework, authors have evaluated the outage performance of a typical secondary receiver (SR) where the corresponding secondary transmitter is considered to be energy constrained. In \cite{varshney2008transporting}, author has studied the fundamental trade-off between information decoding and energy harvesting. However, author has considered that receiver has the ability to decode information and harvest energy from the same received signal, which is very difficult due to practical limitations. To overcome this issue, in \cite{zhang2013mimo} Zhang et. al. have proposed the concept of simultaneous wireless information and power transfer (SWIPT), where receiver harvests energy from the received signal using two methods, namely time switching and power splitting.\par
In relay-based cooperative communication systems, the concept of energy harvesting has also been applied. Authors in \cite{nasir2013relaying}-\cite{ding2014wireless} looked at the performance of a cooperative communication system considering SWIPT enabled relay. However, in both the work, the effect of co-channel interference is ignored. In\cite{krikidis2015relay}, Krikidis has evaluated the outage performance of energy harvesting enabled co-operative communication systems considering spatially random SWIPT enabled relays. In that work, effect of co-channel interference is considered only at relays, not at destination. Further, for simplification, author has ignored the contribution of co-channel interference during energy harvesting by the relays.\par
Considering a non-CR scenario, authors in \cite{zhu2015wireless}-\cite{yu2017outage} have evaluated the performance of cooperative communication system considering a single relay and multiple relays with energy harvesting from co-channel interference and source as well. In those works, authors have considered that the locations of interferers are fixed. However, in practice, communication nodes can be randomly located.\par
In \cite{yang2016outage,xu2016outage}, researchers have investigated the outage probability performance of energy harvesting enabled cooperative cognitive radio network. In those works, authors have ignored the effect of interference from PU. Considering a more general scenario, in \cite{liu2016wireless},\cite{kalamkar2017interference} authors have considered the effect of interference form primary transmitters (PTs). However, in those works, authors have assumed that PU transceivers are clustered within a region. However, PU transceivers in practice can be randomly located. \par
Recently, in \cite{shome2017successful} probability of successful data transmission of energy harvesting co-operative cognitive radio network has been evaluated considering interference from PTs to secondary relay and destination and random locations PU transceivers. In that work, only a single relay has been considered. However, there may be many idle nodes scattered within a two dimensional space, which can be used to relay the signal of source towards the destination. \par
Contributions:\par
Given this background, we have extended our previous work in~\cite{shome2017successful} considering multiple spatially random relays and evaluated successful symbol transmission probability of energy harvesting secondary transmitter.\par
The main objective of this work is to investigate the interesting trade-off between the benefit of energy harvesting and the adverse effect of interference from PTs. The main contribution of this paper are summarized as follows
\par
\begin{itemize}
  \item  In this work, we have considered an energy harvesting co-operative cognitive radio network (CRN), where energy constrained ST tries to transmit its packet to secondary destination (SD) through a secondary relay (SR) which is selected from a group of randomly located SRs within a disc. For this system model, we have evaluated the successful data transmission probability of ST.
  \item We have considered the following relay selection schemes
     \begin{enumerate}
      \item \textbf{Best Composite Channel Based Relay Selection Towards Source (BCCTS).} \par
      We have considered composite channel gain based relay selection criteria by which a relay towards source is selected according to both distance and channel gain. Following this criteria, we have evaluated the successful data communication probability between ST and SD.
       \item \textbf{Best symbol-to-interference ratio (SIR) Towards Source Based Relay Selection (BSTS)}.\par
        In our present work, we derive an analytical approximation of successful data communication probability of energy harvesting ST considering best SIR towards source based relay selection policy. A similar kind of relay selection has been taken into account in \cite{lai2017df} without considering energy harvesting scenario. Moreover, in that paper, locations of the relays are considered to be fixed. In contrast to \cite{lai2017df}, in our present work we have considered a more general case where relays and interferers are randomly located under energy harvesting scenario and evaluated the performance of ST.
       \item \textbf{Best SIR Towards Destination based Relay Selection (BSTD)}.\par
         We have evaluated an analytical upper-bound of probability of successful data communication between ST and SD following BSTD scheme, where the relay, which has the best SIR towards destination is selected to forward the symbol of ST. Even though this relay selection policy is similar to \cite{dhungana2014outage}, the system described in \cite{dhungana2014outage} is interference free and not based on energy harvesting scenario.
     \end{enumerate}
\end{itemize}
Very recently a paper with similar motivation has been published \cite{yan2018outage}, where spatially random relays harvest energy from PTs. Unlike \cite{yan2018outage}, in this present work we have considered that secondary transmitter harvests energy from PTs which makes our system model different from \cite{yan2018outage}. Moreover, in that work authors have not considered the case that secondary destination may employ distributed combining scheme for data decoding purpose, whereas in our work, we have considered that secondary receiver utilizes distributed selection combining scheme to decode the data forwarded by secondary relay.\par
The rest of the paper is organized as follows. In Section II we have described our system model. In Section III and IV we have presented energy harvesting and data transmission policy of ST and SR. In Section V analytical expressions for the probability of successful data transmission of ST via SR has been derived considering the relay selection criterion mentioned above. In section VI numerical results are described and in Section VII we have concluded our work.

Notations:  Throughout the paper, we have used the following notations. $E\left( \cdot\right) $ indicates the expectation operator of random variable, $\parallel \cdot \parallel$ denotes norm of a vector, $\Pr\left(\cdot \right) $ represents the probability of an event and $L(\cdot)$ refers to the Laplace transform operator.

%
%
\section{System Model}
\indent\indent
In this treatise, we consider an energy harvesting enabled co-operative CRN, where there is an ST, an SD, multiple randomly located SRs which are scattered within a circle of radius $R$ centered around ST and multiple primary transceivers which are randomly scattered around the entire 2-D space. We consider that PTs and PRs are randomly scattered according to two independent homogeneous point processes (HPPP) denoted by $\Phi_{pt}$ and $\Phi_{pr}$, respectively. Further, we consider that the geographical densities of PT and PR are $\lambda_{pt}$ and $\lambda_{pr}$, respectively. Without the loss of generality, we consider $\lambda_{pt}=\lambda_{pr}=\lambda_{p}$. Furthermore, the relays are scattered within the circle of radius $R$ according to another HPPP $\Phi_{sr}$. In this present work, we consider that all the communication nodes are equipped with a single omnidirectional antenna. Moreover, it is considered that ST scavenges energy from all PTs while SR and SD do not have any energy issue.
\begin{figure}[h]
\centering
\includegraphics[scale=0.45]{Multi_relay2}
\caption{{A wireless energy harvesting fixed secondary network consisting of one ST, multiple spatially random SR, one SD and multiple randomly located primary transmitter-receiver pairs.  }}
\end{figure}

Since, it is considered that ST and SRs are operating on the same frequency band used by PTs and PRs, both ST and SR have to choose appropriate strategy such that the interference from ST or SR does not degrade the QoS of primary transmission. Like most of the work related to underlay CRN~\cite{duong2012cognitive},  it is assumed that ST controls its transmit power to maintain an interference constraint imposed by PRs. On the other hand, selected SR finds whether it is lying outside any of the guard zones or not, which is defined as the circular area with radius $r_{g}$ around each PR. Otherwise, it is prohibited for data transmission. Moreover, it is assumed that the direct link between ST-SD is severely damaged because of shadowing.
We consider that energy harvesting and data transmission take place at different time slots i.e (i) Energy harvesting slot with duration $a\cdot T$ (ii) ST - SR data transmission slot with duration $(1-a)\cdot T/2$ (iii) SR - SD data transmission slot with duration $(1-a)\cdot T/2$, where $0\leq a \leq 1$. SR-SD data transmission slot will be scheduled if and only if selected SR lies outside all the guard zones.

\par
  In this work, it is considered that the channel between any pair of nodes experiences quasi-static block fading i.e. the channel remains constant for entire time block allotted for energy harvesting and secondary data transmission and varies independently from one block to another. All the channels are assumed to be i.i.d Rayleigh distributed. Hence, the channel power gain turns out to be exponentially distributed.

\par
\section{ Energy harvesting policy of ST}

The energy harvesting policy for ST is described as follows:
 \begin{itemize}
  \item
ST harvests energy from all the PTs in the energy harvesting time slot
\item   Instead of being idle, ST also opportunistically utilizes SR-SD time slot to harvest energy from PTs.
\end{itemize}

Let the transmit power of each PT be $P_{t}$. Let us consider that the total energy harvested by ST is $E_{h}$, then it can be written as follows
\begin{align}
 E_{h} =\eta~P_{t}~a\cdot T\sum_{i\epsilon \Phi_{pt}}  h_{PT_{i}-ST}^{(nes)}~\parallel X_{PT_{i}}^{(nes)} -X_{ST}\parallel^{-\alpha}
 +&\nonumber \\ \eta~P_{t}~\dfrac{1-a}{2}\cdot T\sum_{i\epsilon \Phi_{pt}}  h_{PT_{i}-ST}^{(sr-sd)}~\parallel X_{PT_{i}}^{(sr-sd)} -X_{ST}\parallel^{-\alpha},
 \end{align}
where  $h_{PT_{i}-ST}^{(nes)}$, $X_{PT_{i}}^{(nes)}$ are channel power gains between ST and $i$'th PT at next energy harvesting slot and $i$'th PT during next energy harvesting slot, respectively. Further, $h_{PT_{i}-ST}^{(sr-sd)}$  and $X_{PT_{i}}^{(sr-sd)}$ are channel power gains between ST and $i$'th PT and location of $i$'th PT during SR-SD data transmission slot, respectively.\par
Like \cite{liu2016secure}, it is considered that $P_{st}$ be the threshold transmit power below which transmission of ST does not take place. Let the required transmit energy of ST is  $E_{st}$, then $P_{st}$ and $E_{st}$ are related as $E_{st}=P_{st}\cdot \dfrac{\left( 1-a\right)~T }{2}$. The transmission of ST can be scheduled if and only if $E_{h}\geq E_{st}$. Moreover like \cite{liu2016secure} it is considered that ST is equipped with a super capacitor. In this work, we always consider that $ P_{min}\leq P_{st} \leq P_{max}$, where $P_{min}$ is the minimum required threshold power to activate the energy harvesting circuitry of ST and $P_{max}$ is the maximum transmit power constraint provided for a communication device.\par
Let the probability that ST harvests sufficient energy be $p_{h}$. Then it can be written:
The probability that ST harvests sufficient energy, i.e $p_{h}$ can be written:
\begin{align}
p_{h} &= \Pr\left[ E_{h} \geq \dfrac{1-a}{2}\cdot T \cdot P_{st} \right] \nonumber \\
&=\Pr\left[ K \geq \dfrac{1-a}{2}\cdot \dfrac{P_{st}}{\eta\cdot P_{t}}\right]\nonumber \\
&= \int_{\dfrac{1-a}{2}\cdot \dfrac{P_{st}}{\eta\cdot P_{t} }}^{\infty}~f_{K}\left( x\right)~dx
\end{align}
 Let
\begin{align}
K = k_{1}+k_{2},
\end{align}
where $k_{1}= a\cdot \sum_{i\epsilon \Phi_{pt}}  h_{PT_{i}-ST}^{(nes)}~\parallel X_{PT_{i}}^{(nes)} -X_{ST}\parallel^{-\alpha}$ and $k_{2}=\dfrac{1-a}{2}\cdot \sum_{i\epsilon \Phi_{pt}}  h_{PT_{i}-ST}^{(sr-sd)}~\parallel X_{PT_{i}}^{(sr-sd)} -X_{ST}\parallel^{-\alpha}$. Following \cite{lee2013opportunistic}, it is assumed that energy harvesting process and data transmission process are independent and all the channel gains in current SR-SD data transmission slot to next energy harvesting slot are independent of each other. Then it can be concluded that $k_{1}$ and $k_{2}$ are independent of each other. Therefore the Laplace transform of pdf of $K$ can be written as:
\begin{align}
L_{K}\left(s \right)&=L_{k_{1}+k_{2}}\left(s \right) \nonumber \\
&=L_{k_{1}}\left(s \right) \cdot L_{k_{2}}\left(s \right)
\end{align}
where $L_{k_{1}}\left(s \right)$ and $L_{k_{2}}\left(s \right) $ are the Laplace transforms of the pdfs of $k_{1}$ and $k_{2}$. \par
Now $L_{k_{1}}=E\left[ -sk_{1}\right] $, where $E\left[ \cdot\right] $ is expectation operator. Now further deduction leads to:
\begin{align}
L_{k_{1}} &=E\left[ -sk_{1}\right]\nonumber\\
&=E_{j\in\Phi_{pt}}\prod_{j\in \Phi_{pt}}\left[\int_{0}^{\infty} \exp\left[-\left( as\parallel X_{ST}-X_{PR_{j}}\parallel^{-\alpha}+1\right) h \right]dh \right] \nonumber\\
&\eqa \exp\left[ -2\pi\lambda_{p}\int_{0}^{\infty}\left(1-\dfrac{1}{asx^{-\alpha}+1} \right)x~dx \right]\nonumber\\
&=\exp\left[ -2\pi\lambda_{p}\int_{0}^{\infty}\dfrac{asx}{as+x^{\alpha}}dx\right]
\end{align}

After some mathematical operations we get:
\begin{align}
L_{k_{1}}\left( s\right)= \exp\left(-\lambda_{p}~\pi~\Gamma\left( 1+\dfrac{2}{\alpha}\right)~ \Gamma\left( 1-\dfrac{2}{\alpha}\right)~\left(a\cdot s\right) ^{2/\alpha} \right)
\end{align}
and

Similarly  $L_{k_{2}}\left( s\right) $ can be obtained as:
\begin{align}
L_{k_{2}}\left( s\right) = \exp\left(-\lambda_{p}~\pi~\Gamma\left( 1+\dfrac{2}{\alpha}\right) \Gamma\left( 1-\dfrac{2}{\alpha}\right)\left(\dfrac{1-a}{2}\cdot s\right) ^{2/\alpha} \right)
\end{align}
From (5), (6) we get:
\begin{align}
\begin{split}
L_{K}\left(s \right) =\exp\left(-\lambda_{p}~\pi~\Gamma\left( 1+\dfrac{2}{\alpha}\right)  \Gamma\left( 1-\dfrac{2}{\alpha}\right) ~~~~~~~\right. \\ \left. \left\lbrace \left(\dfrac{1-a}{2}\right) ^{2/\alpha} + \left( a\right) ^{2/\alpha} \right\rbrace\cdot s^{2/\alpha}  \right)
\end{split}
\end{align}
 Inverse Laplace transform of $L_{K}\left( s\right) $ for any arbitrary value of $\alpha$ is intractable. Using Gil-Pelaez theorem the expression for $p_{h}$ is obtained as:
\begin{align}
p_{h}=\frac{1}{2}+\frac{1}{\pi}\cdot \int_{0}^{\infty}\dfrac{Im\left[e^{-jw~\sigma}\cdot \Xi_{1}^{*}\left(w \right)  \right] }{w}~dw ,
\end{align}
\par
where $\sigma=\frac{(1-a)\cdot P_{st}}{2\cdot a~\eta~P_{t}}$.
\par


\section{Data Transmission Policy of ST and SR}
 Since we have considered cognitive underlay scenario, ST needs to adopt power control mechanism, so that the Quality of service (QoS) at each PR does not degrade significantly. If ST lies outside all the guard zones, where each guard zone is defined as a circular area with radius $r_{gz}$ centred around each PR, then only it is allowed for data transmission. A similar restriction is applied to SR as well for forwarding the symbol of ST.

As mentioned in~\cite{xu2016outage,liu2016secure}, we do not consider remaining  energy of ST before current energy harvesting slot. This assumption is valid since in this work we have considered that ST has a supercapacitor to store its transmit energy. Since supercapacitors do not hold charge for a long time due to its self-discharging property, the residual energy remaining in the current energy harvesting block is considered to be negligible.\par
After harvesting sufficient energy, ST selects a particular relay and transmits its packet with power $P_{s}$ to the best selected secondary relay (SR). If the SR successfully decodes the packet transmitted by the ST, it will forward the data with power $P_{st}$. During data transmission, SR tries to evaluate whether it is outside the guard zone or not. If the SR remains outside the guard zone, it is allowed to transmit its data.

\section{Analysis for Successful One Way Symbol Transmission Probability of ST}
In this section, we analytically evaluate the successful data communication probability of ST through SR. We define the  overall successful data communication probability as follows
\begin{align}
P_{succ} = p_{h}~P_{dsucc},
\label{eq8}
\end{align}
where $P_{succ}$ is the successful symbol transmission probability of ST to SD via selected SR.

\subsection{\textbf{Best Composite Channel Based Relay Selection Towards Source} }
In this relay selection scheme ST selects best relay towards considering instantaneous channel gain and distance between ST and SRs.  Prior choosing a relay, ST has to check whether it is residing outside guard zone or not. Let the event that ST remains outside the guard zone be $e_{gz}^{ST}$, then the probability of the event $e_{gz}^{ST}$ is equal to the probability that there is no PR residing within the circle centred around the selected ST with radius $r_{g}$. Let the number of PRs inside the disc $b(ST^{*},r_{g})$ is $N_{0}$ and $N_{0}$ is a Poisson random variable with mean $\lambda_{pr}\pi r_{g}^{2}$. We now write as follows
  \begin{align}
  P\left( e_{gz}^{ST}\right)&= \Pr\left( N_{0} =0\right)\nonumber \\
  &= \exp(-\pi r_{g}^{2}\lambda_{pr}).
  \label{eq130}
  \end{align}\par
  In this case relays send the clear to send signals (CTS) to ST in control channel. ST selects that particular relay whose CTS signal strength is maximum. Mathematically the selection scheme is expressed as:
\begin{align}
b_{ST-SR}= \max_{j\in \Phi_{sr}}\left( h_{ST-SR_{j}}\parallel X_{ST}-X_{SR_{j}}\parallel^{-\alpha}\right)
\end{align}

Let the SIR at the selected relay $\gamma_{SR_{b}}= \dfrac{P_{st}b_{ST-SR}}{I_{1}}$, where $I_{1}$ is the interference at the selected relay. If $\gamma_{SR_{b}}\geq \gamma_{th}$ where $\gamma_{th}$ is the predefined SIR threshold, then selected SR successfully decodes the signal of ST. After successfully decoding the packet of ST, the selected SR forwards the packet of ST to the SD. Let the event that SR remains outside the guard zone be $e_{gz}^{SR}$, then the probability of the event $e_{gz}^{SR}$ can be found as:
  \begin{align}
  P\left( e_{gz}^{SR}\right)&= \exp(-\pi r_{g}^{2}\lambda_{pr}).
  \label{eq131}
  \end{align}
The overall successful symbol transmission probability of ST for this relay selection scheme is defined as:

 \begin{align}
 P_{dsucc}^{b} &= \Pr\left(\gamma_{SR_{b}}\geq \gamma_{th}, e_{gz}^{SR},\gamma_{SD_{b}} \geq \gamma_{th},e_{gz}^{ST} N\geq1\right),\nonumber \\
 &= \Pr\left(\gamma_{SR_{b}}\geq \gamma_{th},\gamma_{SD_{b}} \geq \gamma_{th}\mid N\geq1\right)\Pr\left(N\geq 1 \right)\nonumber \\ ~~~~\times &\Pr\left(e_{gz}^{ST} \right)\times \Pr\left(e_{gz}^{SR} \right),
 \label{eq156}
 \end{align}
 where $\gamma_{SD_{b}}= \frac{P_{st}~\parallel X_{SR_{B}}-X_{SD}\parallel^{-\alpha}}{I_{2}}$ is the SIR at SD and $I_{2}$ is the interference at SD form PTs. $I_{2}$ can be expressed as $I_{2}=P_{t}~\sum_{i\epsilon \Phi_{pt}}  h_{PT_{i}-SD}~\parallel X_{PT_{i}} -X_{SD}\parallel^{-\alpha}$, where $h_{PT_{i}-SD}$ is the channel gain between ith PT and SD.\par
 Let us consider $\Upsilon = \Pr\left(\gamma_{SR_{b}}\geq \gamma_{th},\gamma_{SD_{b}} \geq \gamma_{th},\mid N\geq1\right)$, then we can further write:
 \begin{align}
 \Upsilon &=\Pr\left(\gamma_{SR_{b}}\geq \gamma_{th},\gamma_{SD_{b}} \geq \gamma_{th}\mid N\geq1\right),\nonumber \\
 &= \dfrac{\Pr\left( \gamma_{SR_{B}}\geq \gamma_{th}, \gamma_{SD}\geq \gamma_{th}\right) }{\Pr\left( N\geq 1\right) }\nonumber \\
 & - \dfrac{\Pr\left( \gamma_{SR_{B}}\geq \gamma_{th},\gamma_{SD_{B}}\geq \gamma_{th}\mid N=0\right)\times \Pr\left(N=0 \right)  }{\Pr\left( N\geq1\right) },\nonumber \\
&\eqa \dfrac{\Pr\left( \gamma_{SR_{B}}\geq \gamma_{th}, \gamma_{SD}\geq \gamma_{th}\right) }{\Pr\left( N\geq 1\right) }, \nonumber \\
&\eqb \dfrac{\Pr\left( \gamma_{SR_{B}}\geq \gamma_{th}\right) \Pr\left(  \gamma_{SD}\geq \gamma_{th}\right) }{\Pr\left( N\geq 1\right) }, \nonumber \\
&=  \dfrac{\Pr\left( \gamma_{SR_{B}}\geq \gamma_{th}\right) \Pr\left(  \gamma_{SD}\geq \gamma_{th}\right) }{1-\exp\left( -\pi\lambda_{sr}R^{2}\right)  },
 \end{align}
 where step (a) is done following the fact that, if there is no relay within the disc, no signal will reach at SD from ST since the direct link between ST and SD is unavailable. Hence, $\Pr\left( \gamma_{SR_{B}}\geq \gamma_{th},\gamma_{SD_{B}}\geq \gamma_{th}\mid N=0\right)=0$. Moreover, step (b) is done following the assumption that SIR both at selected SR and SD are independent of each other. \par
 Let $\Psi_{3}= \Pr\left(  \gamma_{SR_{b}} \geq \gamma_{th}\right)$, then it can be written:
 \begin{align}
\Psi_{3}&= \Pr\left[ P_{st}\max_{j\in \Phi_{sr}}\left( h_{ST-SR_{j}}\parallel X_{ST}-X_{SR_{j}}\parallel^{-\alpha}\right) \geq \gamma_{th}I_{1}\right], \nonumber \\
&=1-\left\lbrace \Psi_{31} \right\rbrace,
 \label{eq157}
 \end{align}
where
 \begin{align}
\Psi_{31}=  \Pr\left[ b_{1} \leq \dfrac{\gamma_{th}I_{1}}{P_{st}}\right] .\nonumber
 \end{align}
 and
 \vspace{-3 em}
 The upper-bound of $\Psi_{31}$ can be found as

\begin{align}
\Psi_{31}
 &\leqb \exp\left( -2\pi\lambda_{sr}\int_{0}^{R} E_{I_{1}}\left( \exp\left( -\frac{\gamma_{th}I_{1}l^{\alpha}}{P_{st}}\right)\right) l~dl\right), \nonumber \\
 &\eqc\exp\left(-2\pi\lambda_{sr}\int_{0}^{R} \exp\left(-2\pi\lambda_{p} \int_{0}^{\infty}\dfrac{\dfrac{\gamma_{th}P_{t}l^{\alpha}x}{P_{st}}}{x^{\alpha}+\dfrac{\gamma_{th}P_{t}l^{\alpha}}{P_{st}}}~dx\right) l~dl\right),
    \label{eq158}
  \end{align}

  \textit{Proof:} Proof has been provided in APPENDIX A.

\par
Let $\Psi_{4}=\Pr\left(  \gamma_{SD_{b}} \geq \gamma_{th}\right)$. To evaluate $\Psi_{4}$, we assume that distance between SR and SD is large enough that, the distance between selected relay and SD can be approximated as $d_{ST-SD}$. Then it cam be written:
\begin{align}
\Psi_{4}&\approx\Pr\left(  \dfrac{P_{st}d_{ST-SD}^{-\alpha}h_{SR-SD}}{I_{2}} \geq \gamma_{th} \right)\nonumber \\
&\eqa \exp\left(-2\pi\lambda_{p}\int_{0}^{\infty}\left(1-\frac{1}{1+\dfrac{c^{\alpha}\gamma_{th}P_{t}d_{ST-SD}^{\alpha}}{P_{st}}} \right)c~ dc  \right),
\label{eq162}
\end{align}
where (a) is done using the basic definition of Laplace transform of interference. For $\alpha$=4, $\Psi_{4}$ becomes:
\begin{align}
\Psi_{4}&\approx \exp\left( -\dfrac{\pi^2}{2}\lambda_{p}\sqrt{\dfrac{\gamma_{th}P_{t}d_{ST-SD}^{4}}{P_{st}}}\right).
\label{eq163}
\end{align}
The overall $P_{succ}$ can be found as:
\begin{align}
P_{succ}= p_{h}\Psi_{3}\Psi_{4}\times \left( 1-\exp\left(-\pi\lambda_{p}R^{2} \right) \right).
\end{align}
\subsubsection{\textbf{BCCTS in Presence of Direct link between ST and SD} }\par
In this subsection, coverage probability of ST considering the existence of a direct link between ST and SD under BCCTS scheme. The selected relay forwards the symbol of ST after successfully decoding the symbol of ST. Further, through ST-SD link signal reaches at SD as well. ST employs distributed selection combining scheme to decode the symbol transmitted by ST. If selected relay fails to decode the symbol of ST, SD becomes unable to utilize the distributed selection combining scheme. Here, we consider that ST and SR (if and only if there is non zero relays in a set, which consists of the relays that decodes the symbol of ST correctly and the circular area with radius R is not empty) are allowed to transmit data if and only if both of them are outside guard zone.\par

In presence direct link between ST  and SD, successful symbol transmission probability of ST $P_{succ}$ for BCCTS scheme can be written as
\begin{align}
P_{dsucc}^{dir}&=  \Pr\left( \max\left( \gamma_{ST-SD}, \gamma_{SD_{b}}\right) \geq \gamma_{th}, \gamma_{SR_{b}}\geq \gamma_{th},e_{gz}^{ST}, e_{gz}^{SR},N \geq 1 \right)\nonumber \\
&+ \Pr\left(\gamma_{ST-SD}\geq \gamma_{th} , \gamma_{SR_{b}}\leq \gamma_{th}, e_{gz}^{ST},  N\geq 1\right) \nonumber \\
&+\Pr\left(\gamma_{ST-SD}\geq \gamma_{th},  e_{gz}^{ST},  N=0\right)\nonumber \\
&= \left[1- \Pr\left(\gamma_{ST-SD}\leq \gamma_{th} \right)\Pr\left(\gamma_{SD_{b}}\leq \gamma_{th}\mid N\geq 1 \right)  \right]\\ 
&\times \Pr\left( \gamma_{SR_{b}}\geq \gamma_{th}\mid N \geq 1\right)\Pr\left(N\geq 1 \right)\exp\left(-2\pi\lambda_{p}r_{gz}^2 \right) \nonumber \\
&+\Pr\left( \gamma_{ST-SD}\geq \gamma_{th}\right)\Pr\left( b_{SIR}\leq \gamma_{th}\mid N\geq 1\right)\Pr\left(N\geq 1 \right) \exp\left(-\pi\lambda_{p}r_{gz}^2 \right)\nonumber \\
&+\Pr\left(\gamma_{ST-SD}\geq \gamma_{th}\right)\Pr\left(N=0 \right) \exp\left(-\pi\lambda_{p}r_{gz}^{2} \right),
\label{eq16610}
\end{align}
where $\gamma_{ST-SD}$ is the SIR of the ST-SD link, which can be expressed as:
\begin{align}
\gamma_{ST-SD}=\dfrac{P_{st}h_{ST-SD}d_{ST-SD}^{-\alpha}}{I_{SD}}.
\label{eq1552}
\end{align}
Using the definition of Laplace transform of interference  we get
\begin{align}
\Pr\left(\gamma_{ST-SD}\leq \gamma_{th} \right) = 1-\exp\left(-2\pi\lambda_{p}\int_{0}^{\infty}\left(1-\frac{1}{1+\dfrac{x^{\alpha}\gamma_{th}P_{t}d_{ST-SD}^{\alpha}}{P_{st}}} \right)x~ dx  \right).
\label{eq16620}
\end{align}
It can be realized that, for sufficiently large distance from ST and SD, following  $\Pr\left(\gamma_{SD_{b}} \leq \gamma_{th}\mid N\geq 1\right)$ can approximated as :
\begin{align}
\Pr\left(\gamma_{SD_{b}} \leq \gamma_{th}\mid N\geq 1\right)\approx  1-\exp\left(-2\pi\lambda_{p}\int_{0}^{\infty}\left(1-\frac{1}{1+\dfrac{x^{\alpha}\gamma_{th}P_{t}d_{ST-SD}^{\alpha}}{P_{s}}} \right)x~ dx  \right).
\label{eq16630}
\end{align}
Let $p_{11}= \Pr\left( \gamma_{SR_{b}}\geq \gamma_{th}\mid N \geq 1\right)$, then it can be written as
\begin{align}
p_{11}&= \Pr\left( \gamma_{SR_{b}}\geq \gamma_{th}\right)-\left(1- \Pr\left( b_{SIR}\leq \gamma_{th}\mid N =0\right)\right)\Pr\left(N=0 \right)  \nonumber \\
&\eqa \Pr\left( \gamma_{SR_{b}}\geq \gamma_{th}\right).
\label{eq16640}
\end{align}
Step (a) can be found following the fact that, if there is not any relay within the disc, then it can be written
\begin{align}
\Pr\left( b_{SIR}\leq \gamma_{th}\mid N =0\right) =1.
\end{align} 
Therefore $p_{11}$ can be written as
\begin{align}
p_{11}\geq 1-\exp\left(-2\pi\lambda_{sr}\int_{0}^{R} \exp\left(-2\pi\lambda_{p} \int_{0}^{\infty}\dfrac{\dfrac{\gamma_{th}P_{t}l^{\alpha}x}{P_{st}}}{x^{\alpha}+\dfrac{\gamma_{th}P_{t}l^{\alpha}}{P_{st}}}~dx\right) l~dl\right)
\label{eq16650}
\end{align}
Furthermore, let $p_{12}= \Pr\left( b_{SIR}\leq \gamma_{th}\mid N \geq 1\right)$, then we can write
\begin{align}
p_{12}&= \Pr\left( b_{SIR}\leq \gamma_{th}\right)
-\Pr\left( b_{SIR}\leq \gamma_{th}\mid N =0\right)\Pr\left( N=0\right)\nonumber \\
&= \Pr\left( b_{SIR}\leq \gamma_{th}\right)-\Pr\left(N=0 \right)\nonumber \\
& \leq 1-p_{11}-\exp(-\pi\lambda_{sr}R^2).
\label{eq1666}
\end{align}
Using \eqref{eq16610}, \eqref{eq16620}, \eqref{eq16630},  \eqref{eq16640}, \eqref{eq16650} we get the analytical approximation of $P_{dsucc}^{dir}$ can be obtained.\par
The overall coverage probability of ST can be obtained as
\begin{align}
P_{succ}^{dir} \approx p_{h}\times P_{dsucc}^{dir}.
\end{align}
\vspace{-4 em}
\subsection{\textbf{Best Symbol to Interference Ratio Towards Source-based Relay Selection}}
In this relay selection scheme, SD selects the relay based on the maximum SIR (symbol to interference ratio)-based criteria, where SR with best instantaneous SIR will be selected to forward the symbol of ST. Each SR is equipped with a timer. The stopping time of each timer is inversely proportional to the instantaneous SIR of each relay. Hence, the relay, whose timer expires first has the highest instantaneous SIR and this relay will be selected for transmission to the SD. The relay selection policy can be expressed as follows
 \begin{align}
 b_{SIR}&=\max_{j\in \Phi_{SR}} \left( P_{st}\dfrac{h_{ST-SR_{j}}\parallel X_{ST}-X_{SR_{j}} \parallel ^{-\alpha}}{I_{j}}\right)  \nonumber\\
 &= P_{st}\max_{j\in \Phi_{SR}} \left( \dfrac{h_{ST-SR_{j}}\parallel X_{ST}-X_{SR_{j}} \parallel ^{-\alpha}}{I_{j}}\right),
\label{eq164}
 \end{align}
where $h_{ST-SR_{j}}$ is the channel gain between ST and the $j$'th  relay, $X_{SR_{j}}$ is the location of the $j$'th relay and $I_{j}$ is the interference from PTs at the $j$'th relay. The successful packet transmission of ST for this relay selection scheme is defined as
  \begin{align}
 P_{dsucc}^{bsir} &= \Pr\left(b_{SIR}\geq \gamma_{th},\gamma_{SD_{b_{sir}}} \geq \gamma_{th}, N\geq1, e_{gz}^{ST},e_{gz}^{SR}\right) \nonumber\\
 &= \Pr\left(b_{SIR}\geq \gamma_{th},\gamma_{SD_{b_{sir}}} \geq \gamma_{th}\mid N\geq1\right)~~\nonumber \\&\Pr\left(N\geq 1 \right) \times \Pr\left(e_{gz}^{ST} \right)\times \Pr\left(e_{gz}^{SR} \right),
 \label{eq165}
 \end{align}
where $\gamma_{SD_{b_{sir}}}$ is the SIR of the SD. Now, $\gamma_{SD_{b_{sir}}}$ is defined as follows
\begin{align}
\gamma_{SD_{b_{sir}}}= \dfrac{P_{st}h_{SR_{b}-SD}\parallel X_{SR_{b}}-X_{SD}\parallel^{-\alpha}}{I_{SD}},
\label{eq166}
\end{align}
where $h_{SR_{b}-SD}$ is the channel gain between the selected relay and the SD and $X_{SR_{b}}$ is the location of relay with best the SIR. Let $\Upsilon_{1}=\Pr\left(b_{SIR}\geq \gamma_{th},\gamma_{SD_{b_{sir}}} \geq \gamma_{th}\mid N\geq1\right)$, then we can write as
 \begin{align}
 \Upsilon_{1} &=\Pr\left(b_{SIR}\geq \gamma_{th},\gamma_{SD_{b_{sir}}} \geq \gamma_{th}\mid N\geq1\right),\nonumber \\
 &= \dfrac{\Pr\left( b_{SIR}\geq \gamma_{th}, \gamma_{SD_{bsir}}\geq \gamma_{th}\right) }{\Pr\left( N\geq 1\right) }\nonumber \\
 & - \dfrac{\Pr\left( b_{SIR}\geq \gamma_{th},\gamma_{SD_{bsir}}\geq \gamma_{th}\mid N=0\right)\times \Pr\left(N=0 \right)  }{\Pr\left( N\geq1\right) },\nonumber \\
&\eqa \dfrac{\Pr\left( b_{SIR}\geq \gamma_{th}, \gamma_{SD_{bsir}}\geq \gamma_{th}\right) }{\Pr\left( N\geq 1\right) }, \nonumber \\
&\eqb \dfrac{\Pr\left( b_{SIR}\geq \gamma_{th}\right) \Pr\left(  \gamma_{SD_{bsir}}\geq \gamma_{th}\right) }{\Pr\left( N\geq 1\right) }, \nonumber \\
&=  \dfrac{\Pr\left( b_{SIR}\geq \gamma_{th}\right) \Pr\left(  \gamma_{SD_{bsir}}\geq \gamma_{th}\right) }{1-\exp\left( -\pi\lambda_{sr}R^{2}\right)  }.
\label{eq167}
 \end{align}
Let $\Omega=\Pr\left( b_{SIR}\geq \gamma_{th}\right) $, then we can write as
\begin{align}
\Omega&=\Pr\left(P_{s}\max_{j\in \Phi_{SR}} \left( \dfrac{h_{ST-SR_{j}}\parallel X_{ST}-X_{SR_{j}} \parallel ^{-\alpha}}{I_{j}}\right)\geq \gamma_{th} \right), \nonumber\\
&=1-\Pr\left(P_{s}\times \max_{j\in \Phi_{SR}} \left( \dfrac{h_{ST-SR_{j}}\parallel X_{ST}-X_{SR_{j}} \parallel ^{-\alpha}}{I_{j}}\right)\leq \gamma_{th} \right),\nonumber \\
&=1- \Omega_{1},
\label{eq167}
\end{align}
where $\Omega_{1}=\Pr\left(P_{st}  \max_{j\in \Phi_{SR}} \left( \dfrac{h_{ST-SR_{j}}\parallel X_{ST}-X_{SR_{j}} \parallel ^{-\alpha}}{I_{j}}\right)\leq \gamma_{th} \right)$.
Expression for $\Omega_{1}$ can be found as:
\begin{align}
\Omega_{1}
 &= \exp\left(-2\pi\lambda_{sr}\int_{0}^{R} \exp\left(-2\pi\lambda_{p}\int_{0}^{\infty} \dfrac{x~dx}{1+x^{\alpha}\left(\dfrac{\gamma_{th}P_{t}r^{\alpha}}{P_{s}} \right)^{-1} }\right)r~dr\right),
\label{eq168}
\end{align}
\textit{Proof:} Proof has been provided in APPENDIX B.\par
Let us assume that $\varphi=\Pr\left( \gamma_{SD_{b_{sir}}}\geq \gamma_{th}\right) $. In this case also, we assume that distance between ST and SD is sufficiently large and the distance between the selected relay and the SD can be approximated as $d_{ST-SD}$. Then, the approximated expression for $\varphi$ can be obtained as follows
\begin{align}
\varphi &\approx \Pr\left(  \dfrac{P_{st}d_{ST-SD}^{-\alpha}h_{SR-SD}}{I_{SD}} \geq \gamma_{th} \right)\nonumber \\
&= \exp\left(-2\pi\lambda_{p}\int_{0}^{\infty}\left(1-\frac{1}{1+\dfrac{c^{\alpha}\gamma_{th}P_{t}d_{ST-SD}^{\alpha}}{P_{st}}} \right)c~ dc  \right).
\end{align}
For $\alpha=4$, the $\varphi$ can be approximated as
\begin{align}
\varphi &\approx \exp\left( -\dfrac{\pi^2}{2}\lambda_{p}\sqrt{\dfrac{\gamma_{th}P_{t}d_{ST-SD}^{4}}{P_{st}}}\right).
\end{align}

\subsubsection{\textbf{BSIRTS in Presence of Direct link between ST and SD} }\par
In presence of direct link between ST and SD, coverage probability of ST can be defined as
\begin{align}
P_{dsucc}^{dir}&=  \Pr\left( \max\left( \gamma_{ST-SD}, \gamma_{SD_{b_{sir}}}\right) \geq \gamma_{th}, b_{SIR}\geq \gamma_{th},e_{gz}^{ST}, e_{gz}^{SR},N \geq 1 \right)\nonumber \\
&+ \Pr\left(\gamma_{ST-SD}\geq \gamma_{th} , b_{SIR}\leq \gamma_{th}, e_{gz}^{ST},  N\geq 1\right) \nonumber \\
&+\Pr\left(\gamma_{ST-SD}\geq \gamma_{th},  e_{gz}^{ST},  N=0\right)\nonumber \\
&= \left[1- \Pr\left(\gamma_{ST-SD}\leq \gamma_{th} \right)\Pr\left(\gamma_{SD_{bsir}}\leq \gamma_{th}\mid N\geq 1 \right)  \right]\\
&\times \Pr\left( b_{SIR}\geq \gamma_{th}\mid N \geq 1\right)\Pr\left(N\geq 1 \right)\exp\left(-2\pi\lambda_{p}r_{gz}^2 \right) \nonumber \\
&+\Pr\left( \gamma_{ST-SD}\geq \gamma_{th}\right)\Pr\left( b_{SIR}\leq \gamma_{th}\mid N\geq 1\right)\Pr\left(N\geq 1 \right) \exp\left(-\pi\lambda_{p}r_{gz}^2 \right)\nonumber \\
&+\Pr\left(\gamma_{ST-SD}\geq \gamma_{th}\right)\Pr\left(N=0 \right) \exp\left(-\pi\lambda_{p}r_{gz}^{2} \right).
\label{eq1661}
\end{align}
From \eqref{eq16620} we get
\begin{align}
\Pr\left(\gamma_{ST-SD}\leq \gamma_{th} \right) = 1-\exp\left(-2\pi\lambda_{p}\int_{0}^{\infty}\left(1-\frac{1}{1+\dfrac{x^{\alpha}\gamma_{th}P_{t}d_{ST-SD}^{\alpha}}{P_{s}}} \right)x~ dx  \right).
\label{eq1662}
\end{align}
For sufficiently large distance from ST and SD, $\Pr\left(\gamma_{SD_{bsir}} \leq \gamma_{th}\mid N\geq 1\right)$ can approximated as :
\begin{align}
\Pr\left(\gamma_{SD_{bsir}} \leq \gamma_{th}\mid N\geq 1\right)\approx  1-\exp\left(-2\pi\lambda_{p}\int_{0}^{\infty}\left(1-\frac{1}{1+\dfrac{x^{\alpha}\gamma_{th}P_{t}d_{ST-SD}^{\alpha}}{P_{s}}} \right)x~ dx  \right).
\label{eq1663}
\end{align}
Let $p_{22}= \Pr\left( b_{SIR}\geq \gamma_{th}\mid N \geq 1\right)$, then it can be written as
\begin{align}
p_{22}&= \Pr\left( b_{SIR}\geq \gamma_{th}\right)-\left(1- \Pr\left( b_{SIR}\leq \gamma_{th}\mid N =0\right)\right)\Pr\left(N=0 \right)  \nonumber \\
&\eqa \Pr\left( b_{SIR}\geq \gamma_{th}\right).
\label{eq1664}
\end{align}
Step (a) is performed following the fact that, if there is not a single relay within the disc, then it can be written
\begin{align}
\Pr\left( b_{SIR}\leq \gamma_{th}\mid N =0\right) =1.
\end{align}

 Therefore it can be written 
\begin{align}
p_{22}=1-\exp\left(-2\pi\lambda_{sr}\int_{0}^{R} \exp\left(-2\pi\lambda_{p}\int_{0}^{\infty} \dfrac{x~dx}{1+x^{\alpha}\left(\dfrac{\gamma_{th}P_{t}r^{\alpha}}{P_{s}} \right)^{-1} }\right)r~dr\right).
\label{eq1665}
\end{align}
Furthermore, let $p_{32}= \Pr\left( b_{SIR}\leq \gamma_{th}\mid N \geq 1\right)$, then we can write:
\begin{align}
p_{32}&= \Pr\left( b_{SIR}\leq \gamma_{th}\right)
-\Pr\left( b_{SIR}\leq \gamma_{th}\mid N =0\right)\Pr\left( N=0\right)\nonumber \\
&= \Pr\left( b_{SIR}\leq \gamma_{th}\right)-\Pr\left(N=0 \right)\nonumber \\
& =1- p_{22}-\exp(-\pi\lambda_{sr}R^2).
\label{eq1666}
\end{align}
Using \eqref{eq1661}, \eqref{eq1662}, \eqref{eq1663},  \eqref{eq1664}, \eqref{eq1665} we get the analytical approximation of $P_{dsucc}^{dir}$ can be obtained.\par
The overall coverage probability of ST can be obtained as
\begin{align}
P_{succ}^{dir} \approx p_{h}\times P_{dsucc}^{dir}.
\end{align}


\subsection{\textbf{Best SIR Towards Destination based Relay Selection}}
In this subsection, a different relay selection criteria has been considered, where SD will select a particular relay from a set of relays, which have correctly decoded the symbol transmitted by the ST. The SD will select a particular relay from this set according to the best SIR-based selection rule. As discussed earlier, SD will choose a suitable relay from a set of relays, which have successfully decoded the symbol transmitted by ST, hence it is necessary to find out the probability that a typical relay correctly decodes the symbol transmitted by ST. Let the probability that a typical relay i.e the $i^{th}$ relay successfully decodes the symbol transmitted by ST and ST is outside guard zone be $\Delta$, then it can be written
\begin{align}
\Delta &= \Pr\left( P_{s}~h_{ST-SR_{i}} \parallel X_{ST}-X_{SR_{i}}\parallel^{-\alpha}\dfrac{1}{I_{SR_{i}}}\geq \gamma_{th},e_{gz}^{ST}\right) \nonumber\\
&= E_{I_{SR}}\left( \int_{0}^{R}\exp\left( -\dfrac{I_{SR}\gamma_{th}r^{\alpha}}{P_{s}}\right) \dfrac{2r}{R^2}~dr\right)~\exp\left(-\pi\lambda_{p}r_{gz}^{2} \right) \nonumber\\
&\eqa \int_{0}^{R} ~\exp\left( -2\pi\lambda_{p}\int_{0}^{\infty}\dfrac{x}{1+\left( \dfrac{\gamma_{th}r^{\alpha}P_{t}}{P_{s}}\right)^{-1} x^{\alpha}}~dx\right)~\dfrac{2r}{R^{2}} ~dr~ \exp\left(-\pi\lambda_{p}r_{gz}^{2} \right),
\label{eq166}
\end{align}
where $X_{SR_{i}}$ is the location of the $i^{th}$ relay, $I_{SR_{i}}$ is the interference at the $i^{th}$ relay and $h_{ST-SR_{i}}$ is the channel power gain between ST and $i^{th}$ relay. Further, the first of step of~\eqref{eq166} has been obtained invoking the Laplace transform of interference. For $\alpha=4$, \eqref{eq166} becomes
  \begin{align}
  \Delta &= \int_{0}^{R^2} \exp\left(-\dfrac{\pi^2}{2}\lambda_{p} \left( \dfrac{\gamma_{th}P_{t}}{P_{s}}\right)^{0.5}z \right) ~dz ~\exp\left(-\pi\lambda_{p}r_{gz}^{2} \right)\nonumber\\
  &= \dfrac{2}{\left(R^{2}\pi^2\lambda_{p}\left( \dfrac{\gamma_{th}P_{t}}{P_{s}}\right)^{0.5}  \right) }\times \left[1-\exp\left(-\dfrac{\pi^2}{2}\lambda_{p}\left(\dfrac{\gamma_{th}P_{t}}{P_{s}} \right)^{0.5}R^{2}  \right)  \right]~\exp\left(-\pi\lambda_{p}r_{gz}^{2} \right).
\label{eq167}
  \end{align}
Following the thinning process of PPP, the effective density of relays, which have correctly decoded the symbol of ST can be written as $\lambda_{sr_{eff}}= \Delta\lambda_{sr}$. After successfully decoding the symbol of ST, relays send request-to-send (RTS) signal towards the SD in control channel. The SD  measures the received signal power level from the RTS signals. It also measures the interference power level impinging on it from various PTs. With this information, SD identifies the best relay that has the maximum SIR towards it. Then SD sends a clear-to-send (CTS) signal to that particular relay. Other relays will not respond to that CTS signal, since it is not meant for them. If the selected relay is out of all the guard zones centred around PRs, it will be allowed to forward the symbol of ST, otherwise not. \par
Let the probability of successfully decoding the message forwarded by the selected SR be $P_{dsucc}^{sd}$. Then the probability of successful decoding at the destination can be written as
\vspace{-1 em}
\begin{align}
P_{dsucc}^{sd} & = \Pr\left(\max_{i\in \Phi_{sr}^{0}} \dfrac{P_{st}h_{SR_{i}-SD} f\left( r_{i}\right)^{-\alpha} }{I_{SD}}\geq \gamma_{th}, e_{gz}^{SR} \right), \nonumber\\
 &= \Pr\left(\max_{i\in \Phi_{sr}^{0}} \dfrac{P_{st}h_{SR_{i}-SD} f\left( r_{i}\right)^{-\alpha}  }{I_{SD}}\geq \gamma_{th}, e_{gz}^{SR} \right), \nonumber\\
 &= \left( 1-\Pr\left(\max_{i\in \Phi_{sr}^{0}} \dfrac{P_{st}h_{SR_{i}-SD} \left( f\left( r_{i}\right) \right)^{-\alpha}}{I_{SD}}\leq \gamma_{th} \right)\right)  \Pr\left( e_{gz}^{SR}\right),
\label{eq168}
\end{align}
where $f\left( r_{i}\right)= \sqrt{r_{i}^2+d_{ST-SD}^2-2r_{i}d_{ST-SD} ~cos\left( \theta\right) } $ is the distance between the $i^{th}$ SR and the SD and $(r_{i}, \theta)$ is the coordinate of $i^{th}$ relay. Let $\chi \triangleq \Pr\left(\max_{i\in \Phi_{sr}^{0}} \dfrac{P_{st}h_{SR_{i}-SD} \left( f\left( r_{i}\right) \right)^{-\alpha}}{I_{SD}}\leq \gamma_{th} \right)$, then  we evaluate it as
\begin{align}
\chi &= E_{I_{SD},\Phi_{sr}^{0}}\prod_{i\in \Phi_{sr}^{0}} \left(1-\exp\left(-\dfrac{\gamma_{th} I_{2}\left( \sqrt{r_{i}^2+d^2-2r_{i}d ~cos\left( \theta\right) }\right)^{\alpha} }{P_{st}} \right)  \right),\nonumber\\
&\eqa E_{I_{SD}} \exp\left(-\lambda_{sr}\Delta \int_{0}^{2\pi}\int_{0}^{R} \exp\left(-\dfrac{\gamma_{th}\left( \sqrt{r^2+d_{ST-SD}^2-2rd_{ST-SD} ~cos\left( \theta\right) }\right)^{\alpha} I_{2}}{P_{st}} \right) r~dr\right), \nonumber\\
&\leqb \exp\left(-\lambda_{sr}\Delta \int_{0}^{2\pi}\int_{0}^{R} E_{I_{2}}\left( \exp\left(-\dfrac{\gamma_{th}\left( \sqrt{r^2+d_{ST-SD}^2-2rd_{ST-SD} ~cos\left( \theta\right) }\right)^{\alpha} I_{SD}}{P_{st}} \right)\right)  r~dr\right).
\label{eq16992}
\end{align}
The first step of~\eqref{eq16992} is performed using PGFL of PPP, and the second step of~\eqref{eq169}  is obtained following the Jensen's inequality. Using the definition of Laplace transform of interference, \ref{eq16992} can be written as
\begin{align}
\xi &= E_{I_{SD}}\left( \exp\left(-\dfrac{\gamma_{th}\left( \sqrt{r^2+d_{ST-SD}^2-2rd ~cos\left( \theta\right) }\right)^{\alpha} I_{SD}}{P_{st}} \right)\right), \nonumber\\
&\eqa\exp\left( -2\pi\lambda_{p}\int_{0}^{\infty}\dfrac{x}{1+\left( \dfrac{\gamma_{th}\left( \sqrt{r^2+d_{ST-SD}^2-2rd_{ST-SD} ~cos\left( \theta\right) }\right)^{\alpha}P_{t}}{P_{st}}\right)^{-1} x^{\alpha}}~dx\right),
\label{eq170}
\end{align}
where the first step of~\eqref{eq170} is obtained following the definition of Laplace transform of interference. For $\alpha=4$, $\xi$ becomes
  \begin{align}
  \xi = \exp\left(-\dfrac{\pi^2}{2}\lambda_{p} \left( \dfrac{\gamma_{th}P_{t}}{P_{st}}\right)^{0.5} \left(r^2+d_{ST-SD}^2-2rd_{ST-SD} ~cos\left( \theta\right) \right) \right).
\label{eq171}
  \end{align}
  For any arbitrary value of $\alpha$, the expression for $\chi$ can be found using~\eqref{eq169} and~\eqref{eq170}. For $\alpha=4$, the expression for $\xi$ can be found using~\eqref{eq170} and~\eqref{eq171}. The overall successful symbol transmission probability of ST is evaluated as
  \begin{align}
  P_{succ} =  p_{h}~P_{dsucc}^{sd}.
  \end{align}
  \vspace{-4.5 em}
\subsubsection{\textbf{BSTD Scheme with Direct Link Between ST and SD}}
  For BSTD scheme, the coverage probability can be defined as:
  \begin{align}
  P_{dsucc}^{dir} &= \Pr\left(\max\left( \gamma_{ST-SD},\max_{i\in \Phi_{sr}^{0}} \dfrac{P_{st}h_{SR_{i}-SD} f\left( r_{i}\right)^{-\alpha} }{I_{2}}\geq \gamma_{th}\right)\geq \gamma_{th}, e_{gz} , N_{1}\geq 1 \right)\nonumber \\
  &+\Pr\left(\gamma_{ST-SD}\geq \gamma_{th}, N_{1}=0, e_{gz}^{ST} \right) \nonumber \\
&=  \left[ 1-\Pr\left( \gamma_{ST-SD}\leq \gamma_{th}\right) \left( \Pr\left( \max_{i\in \Phi_{sr}^{0}} \dfrac{P_{st}h_{SR_{i}-SD} f\left( r_{i}\right)^{-\alpha} }{I_{2}}\leq \gamma_{th}\right)-\Pr\left(N_{1}=0 \right) \right)  \right]\nonumber \\& \times \exp \left( -\pi\lambda_{p}r_{gz}^{2}\right)+\Pr\left(\gamma_{ST-SD}\geq \gamma_{th} \right)\Pr\left(  N_{1}=0\right)\times \exp \left( -\pi\lambda_{p}r_{gz}^{2}\right).
  \label{eq173}
  \end{align}
 Where $N_{1}$ indicates the number of relays which have successfully decoded the symbol of ST. Using the pmf of Poisson distribution $\Pr\left( N_{1}=0\right) $ can obtained as follows:
 \begin{align}
 \Pr\left( N_{1}=0\right) =\exp(-\pi\lambda_{eff}R^2).
\label{eq174}
  \end{align}
 Moreover, it can be written:
  \begin{align}
 \Pr\left( N_{1}\geq 1\right) =1-\exp(-\pi\lambda_{eff}R^2).
\label{eq175}
 \end{align}
 From \eqref{eq1662}, \eqref{eq171}, \eqref{eq173}, \eqref{eq174}, \eqref{eq175} we get the expression for the upperbound of $P_{dsucc}^{dir}$. \par
 Hence the overall successful data transmission probability for ST to SD via chosen SR can be obtained as:
\begin{align}
P_{succ} \geq p_{h}\times P_{dsucc}^{dir}.
\end{align}
\section{Numerical Results}
We present the numerical results based on our analysis. We consider the following parameters. $T=1$ msec, time fraction $a=0.5$, energy harvesting efficiency $\eta=0.8$ and pathloss exponent $\alpha = 4$ and the radius of the circular region over which relays are located, i.e. $R= 1 $ m . For simulation purpose, $3\times10^{4}$ independent realizations of the entire network have been used.

\begin{figure}[!h]
\centering
\includegraphics[scale=0.6]{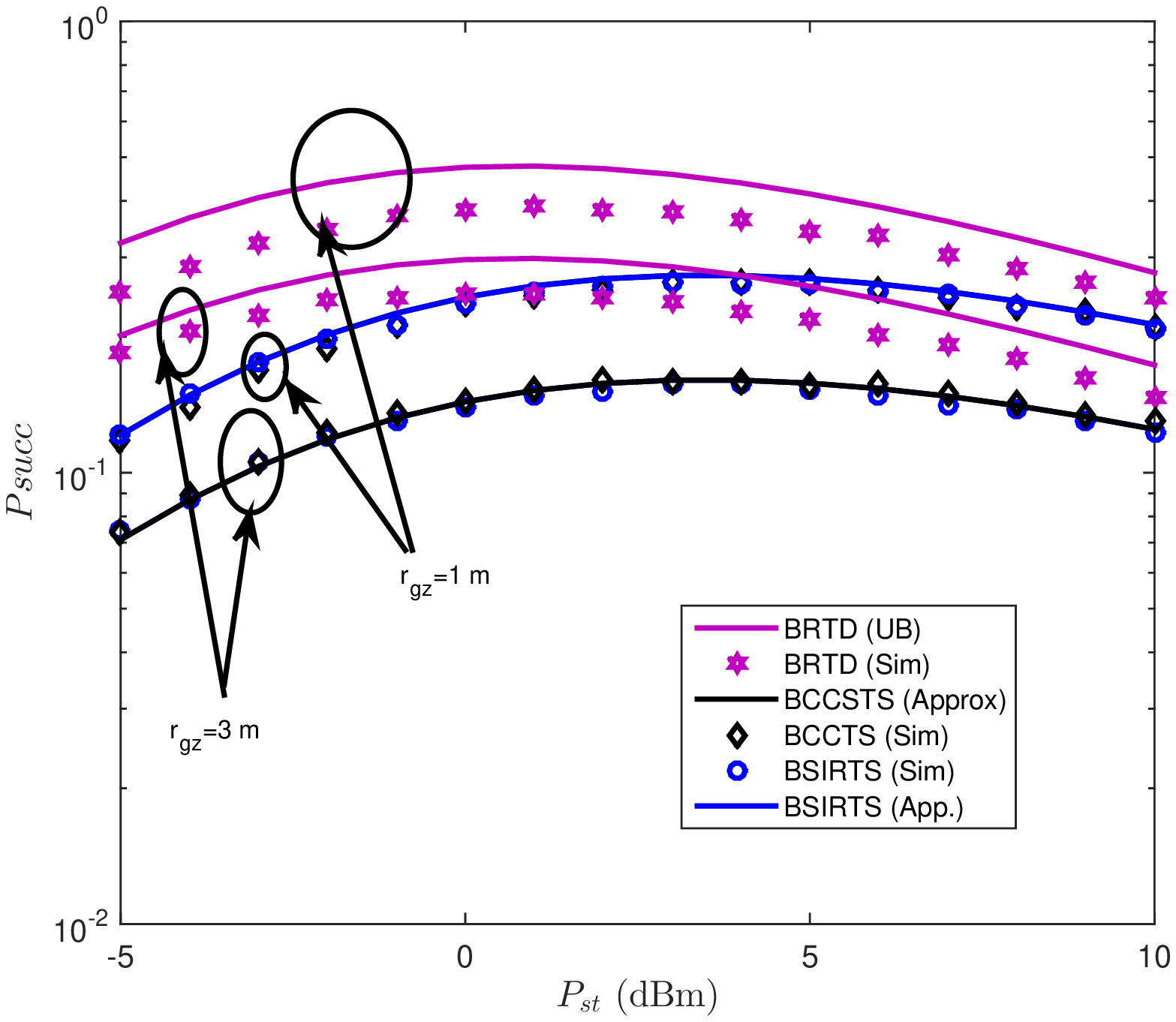}
\caption{{Probability of successful one way data transmission between secondary transmitter and receiver via secondary relay versus required transmit power of secondary transmitter, with $X_{ST}=(0,0)$,  $X_{SD}=(2,0)$, $\gamma_{th}$= -10 dB, $\alpha=4$, $P_{t}= 25$ dBm, $\lambda_{sr}$=1. }}
\label{l1}
\end{figure}
 Fig.~\ref{l1} represents the relationship between the probability of successful data transmission $P_{succ}$ and the required transmit power of ST i.e $P_{st}$. It can be observed that for both NRS  and BSIRTS schemes, if $- 5 \leq P_{st} \leq 0$ dBm (approx), the performance of SD improves, because ST is able to transmit with higher transmit power. However, for $P_{st}\geq 0$ dBm, $P_{succ}$ degrades. The reason behind this is the following. In the aforementioned range, the harvested energy by the ST has to cross higher threshold values. For this reason, $p_{h}$ reduces sharply, consequently $p_{succ}$ is also gradually deteriorated. Moreover, if the radius of the guard zone is increased, selected SR has to follow more restriction on its data transmission towards SD. As a consequence, the overall probability of successful data transmission of ST is deteriorated.

\begin{figure}[!h]
\centering
\includegraphics[scale=0.6]{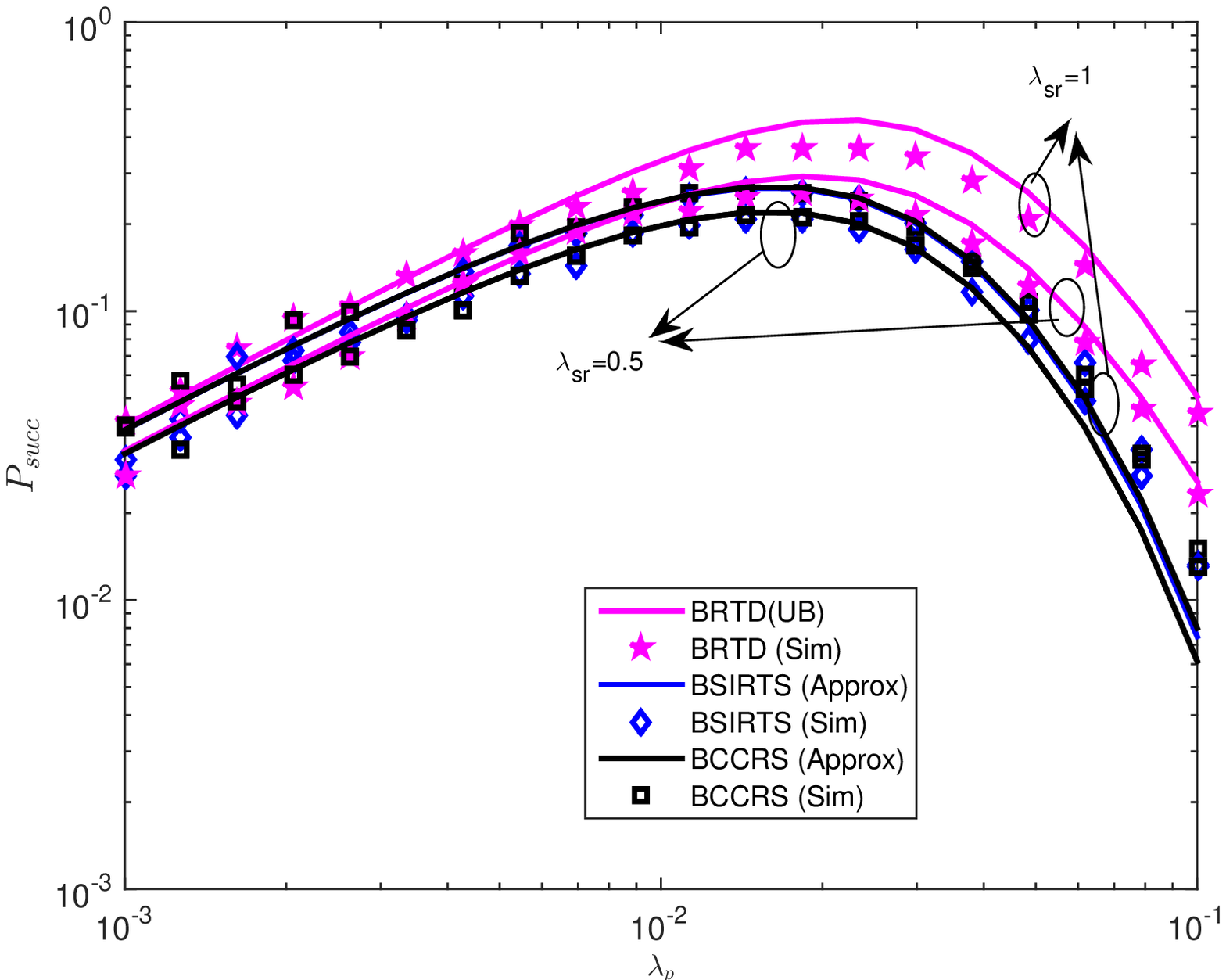}
\caption{{Probability of successful one way data communication between secondary transmitter and receiver via secondary relay versus density of primary users, with $P_{t}=15$ dBm, $P_{st}=-2$ dBm, $X_{ST}=(0,0)$, $X_{SD}=(2,0)$ , $\gamma_{th}$=-10 dB. }}
\label{l2}
\end{figure}

Fig.~\ref{l2} shows the relationship between $p_{succ}$ and the density of PTs i.e $\lambda_{p}$. It can be found that for $10^{-1}\leq\lambda_{p}\leq 10^{-2}$ (approx), with the increase of $\lambda_{p}$, the probability of harvesting sufficient energy to schedule the transmission of ST i.e $p_{h}$ is increased. However, further increase in $\lambda_{p}$ adversely affects the $P_{succ}$, because interference level form PTs to SR and SD increases significantly. As a consequence, $p_{succ}$ is detoriated when $\lambda_{p}\geq 10^{-2}$. Further, it can be observed that if the density of relays is increased, the probability of successful packet transmission of ST improves because of better utilization of relay selection diversity by ST.

\par
\begin{figure}[!h]
\centering
\includegraphics[scale=0.6]{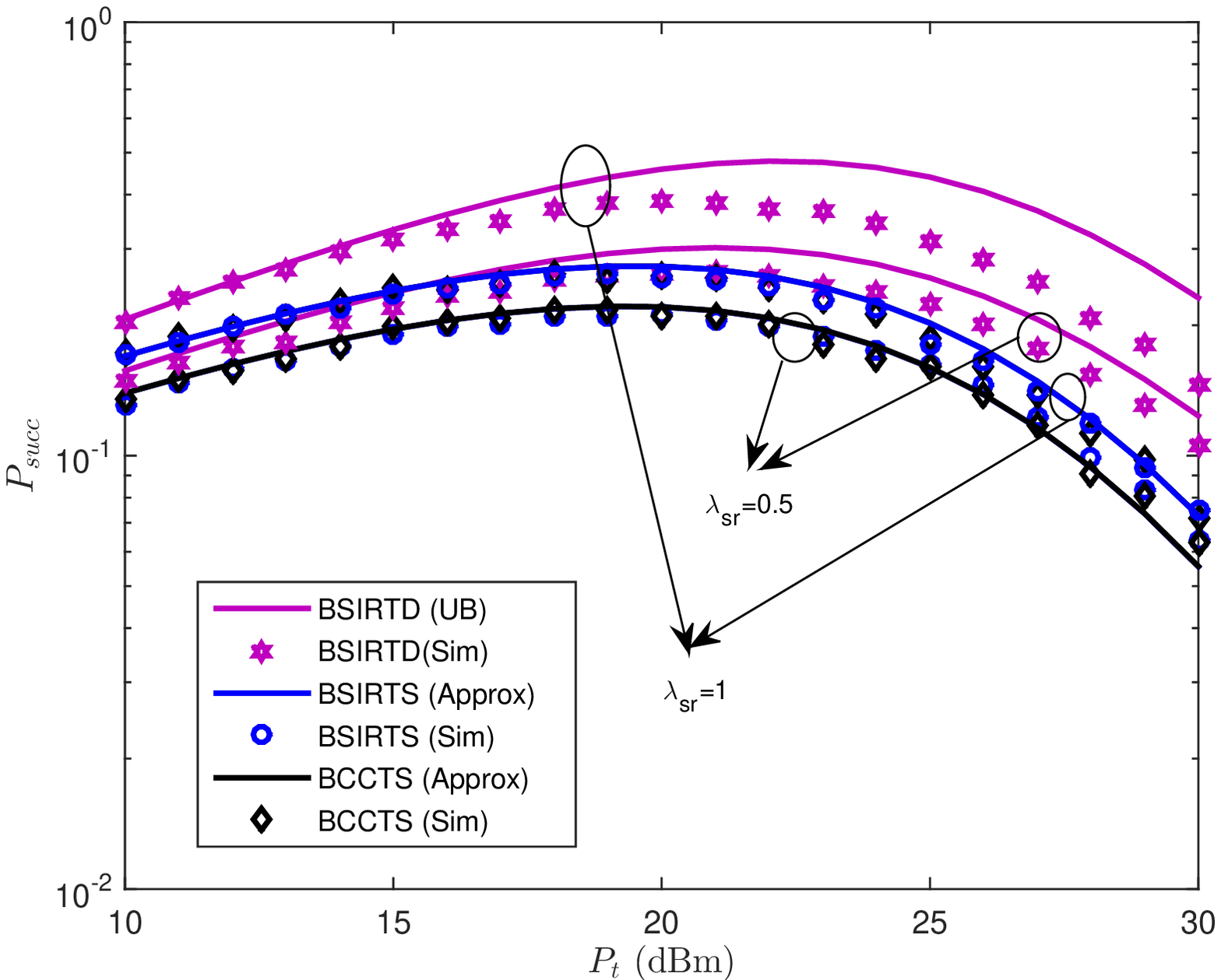}
\caption{{Probability of successful one way data communication between secondary transmitter and receiver via secondary relay versus  transmit power of primary transmitters with  $\gamma_{th}$=-10 dB, $r_{gz}= 1 $ m, $\alpha$=4 and $P_{st}$= -2 dBm. }}
\label{l3}
\end{figure}

Fig.~\ref{l3} shows the relationship between the $P_{succ}$ and the transmit power of PTs i.e $P_{t}$. It can be noticed that for all the relay selection schemes mentioned  throughout the paper, for $ 10\leq P_{t} \leq 23$ dBm, $P_{succ}$ improves, because in that range of $P_{t}$, $p_{h}$ increases quite significantly. However, when $P_{t} \geq 20$ dBm, the $P_{succ}$ is deteriorated, because in that range of $P_{t}$ values, interference from PTs becomes so detrimental that it overturns the befit of energy harvesting from PTs. Further, we can find that, with the increase of the density of SR, overall probability of successful data transmission between ST and SD via selected SR improves according to the reason provided earlier.

\par
\begin{figure}[!h]
\centering
\includegraphics[scale=0.6]{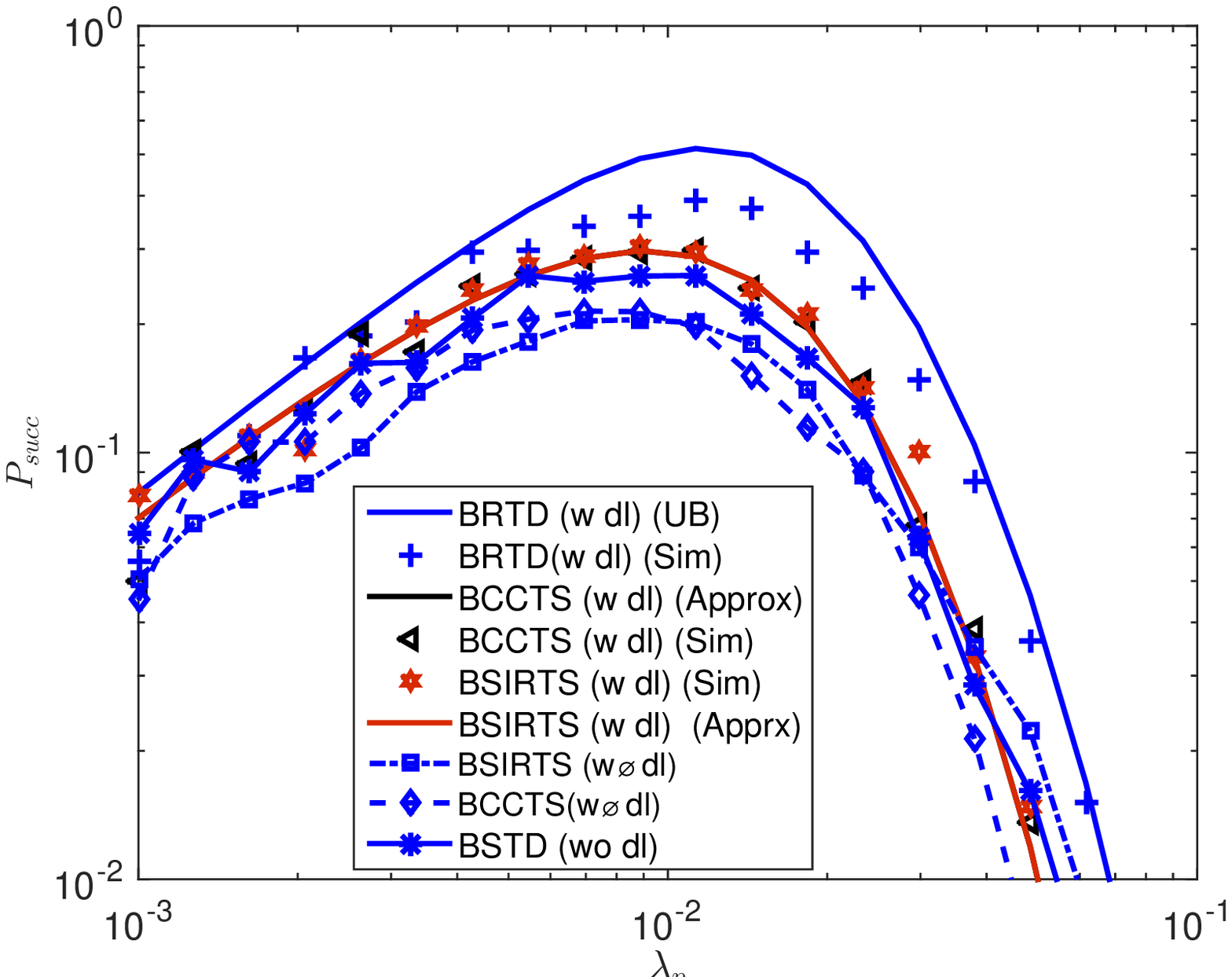}
\caption{{Probability of successful one way data communication between secondary transmitter and receiver via secondary relay versus  transmit power of primary transmitters in presence of direct link between ST and SD with  $\gamma_{th}$=-10 dB,$r_{gz}=2$ m, $\alpha$=4 and $P_{st}$= -2 dBm . }}
\label{l4}
\end{figure}

Fig.~\ref{l4} represents the relationship between $P_{succ}$ and $\lambda_{p}$ considering the direct link between ST and SD. Since, in presence of ST-SD link, SD has the opportunity to employ distributed selection diversity scheme, the $P_{succ}$ in the case of presence of ST-SD link is higher compared to the performance of ST while ST-SD link and is severely damaged due to obstacle or high attenuation due to shadowing.

\section{Conclusion}
\vspace{-1 em}
In this paper, we have evaluated the successful data communication probability, i.e. $P_{succ}$ between ST and SD considering spatially random relays and Poisson field of interferers (PT)under energy harvesting scenario considering various relay selection schemes like NRS, BCCTS, BRTS and BRTD schemes. We have shown the relation between $P_{succ} $ and various parameters like $P_{st}$, $\lambda_{p}$, $\lambda_{sr}$, $P_{t}$ and $r_{gz}$ etc. for each relay selection criteria. The analytical expressions of $P_{succ}$ for each selection criteria have been evaluated using stochastic geometry and validated through simulation work. It is observed that interference from PTs is advantageous for energy harvesting, however after a certain limit interference from PTs has detrimental effect on $P_{succ}$ . We observe that BSTD scheme outperforms all the other schemes stated above.

\appendices

\section{Proof of proposition 1}\label{App A}
\vspace{-4 em}
\begin{align}
\Psi_{31}&= \Pr\left[ \max_{j\in \Phi_{sr}}\left\lbrace h_{ST-SR_{j}}\parallel X_{ST}-X_{SR_{j}}\parallel^{-\alpha}\right\rbrace \leq \dfrac{\gamma_{th}I_{1}}{P_{st}}  \right],  \nonumber \\
&= E_{I_{1}}\left[E_{\Phi_{sr}}\left(\prod_{j\in \Phi_{sr}}\left( 1-\exp\left(- \frac{I_{1}\gamma_{th}\parallel X_{ST}-X_{SR_{j}}\parallel^{\alpha}}{P_{st}}\right) \right)  \right)  \right],   \nonumber \\
&\eqa E_{I_{1}}\left(\exp\left( -2\pi\lambda_{sr}\int_{0}^{R}\exp\left( -\frac{\gamma_{th}I_{1}l^{\alpha}}{P_{st}}\right)l~dl\right)  \right), \nonumber \\
 &\leqb \exp\left( -2\pi\lambda_{sr}\int_{0}^{R} E_{I_{1}}\left( \exp\left( -\frac{\gamma_{th}I_{1}l^{\alpha}}{P_{st}}\right)\right) l~dl\right), \nonumber \\
 &\eqc\exp\left(-2\pi\lambda_{sr}\int_{0}^{R} \exp\left(-2\pi\lambda_{p} \int_{0}^{\infty}\dfrac{\dfrac{\gamma_{th}P_{t}l^{\alpha}x}{P_{st}}}{x^{\alpha}+\dfrac{\gamma_{th}P_{t}l^{\alpha}}{P_{st}}}~dx\right) l~dl\right).
    \label{eq158}
  \end{align}
  where (a) is done using probability generating functional of PPP \cite{chiu2013stochastic}, (b) is done following Jensen's inequality and (c) is done using the definition of Laplace transform of interference.\par
For $\alpha=4$, $\Psi_{31}$ becomes:
\begin{align}
\Psi_{31} &\leq \exp\left( -2\pi\lambda_{sr}\int_{0}^{R}\exp\left(-\dfrac{\pi^{2}}{2}\lambda_{p}\sqrt{\dfrac{\gamma_{th}P_{t}}{P_{st}}}l^{2} \right) l~dl\right)\nonumber \\
&= \exp\left( -\dfrac{2\lambda_{sr}}{\pi\lambda_{p}\sqrt{\dfrac{\gamma_{th}P_{t}}{P_{st}}}} \left( 1-\exp\left(-\dfrac{\pi^{2}}{2}\lambda_{p}\sqrt{\dfrac{\gamma_{th}P_{t}}{P_{st}}} R^{2}\right) \right)\right).
\label{eq159}
\end{align}

\section{Proof of proposition 2}\label{App B}
\vspace{-4 em}
\begin{align}
\Omega_{1} &=E_{\Phi_{SR}}\left[ \prod_{j\in \Phi_{sr} }\Pr\left( \dfrac{P_{s}h_{ST-SR_{j}}\parallel X_{ST}-X_{SR_{j}}\parallel^{-\alpha}}{I_{j}}\leq \gamma_{th}\right) \right], \nonumber\\
 &=E_{\Phi_{SR}}\left[ \prod_{j\in \Phi_{sr}}E_{I_{j}}\left[1-\exp\left( -\dfrac{\gamma_{th}I_{j}\parallel X_{SR_{j}}\parallel^{\alpha}}{P_{s}}\right)  \right] \right]\nonumber\\
 &\eqa E_{\Phi_{SR}}\left[ \prod_{j\in \Phi_{sr}}\left( 1-\exp\left(-2\pi\lambda_{p}\int_{0}^{\infty} \dfrac{x~dx}{1+x^{\alpha}\left(\dfrac{\gamma_{th}P_{t}r_{j}^{\alpha}}{P_{s}} \right)^{-1} }\right) \right) \right]\nonumber\\
 &\eqb \exp\left(-2\pi\lambda_{sr}\int_{0}^{R} \exp\left(-2\pi\lambda_{p}\int_{0}^{\infty} \dfrac{x~dx}{1+x^{\alpha}\left(\dfrac{\gamma_{th}P_{t}r^{\alpha}}{P_{s}} \right)^{-1} }\right)r~dr\right),
\label{eq168}
\end{align}
where~step(a) is obtained following basic definition of Laplace transform of interference~\cite{haenggi2009stochastic} and step (b) is obtained following the PGFL of PPP \cite{chiu2013stochastic}.
For $\alpha=4$, the $\Omega_{1}$ becomes
\begin{align}
\Omega_{1}=\exp\left(-2\lambda_{sr}\dfrac{\left(1-\exp\left(-\dfrac{\pi^2}{2}R^{2}\left(\dfrac{\gamma_{th}P_{t}}{P_{s}} \right)^{0.5}  \right)  \right) }{\pi\lambda_{p}\left(\dfrac{\gamma_{th}P_{t}}{P_{s}} \right)^{0.5} } \right).
\label{eq169}
\end{align}

\end{document}